\documentclass[sigconf, screen=true, review=false, printacmref=false, printccs=false, printfolios=false]{acmart}
\usepackage{comment}

\usepackage{cleveref}
\usepackage{hyperref}

\usepackage{multirow}
\setcopyright{none}

\usepackage{paralist}

\usepackage{booktabs} 

\usepackage{epstopdf}
\usepackage{comment}
\usepackage{tabularx}
\usepackage{subfigure}
\usepackage{enumitem}

\DeclareMathAlphabet\mathbfcal{OMS}{cmsy}{b}{n}

\usepackage{xcolor}
\definecolor{thedarkblue}{RGB}{0,0,120} 
\definecolor{theblue}{rgb}{0,0.08,0.45} 
\definecolor{mydarkblue}{rgb}{0,0,120} 
\definecolor{darkblue}{rgb}{0,0.08,180}
\colorlet{TufteRed}{red!80!black}

\definecolor{theblue}{RGB}{0,0,180}
\colorlet{thered}{TufteRed}

\usepackage{microtype}
\usepackage{balance}

\usepackage{amsmath,amssymb,amsthm}

\usepackage{tikz}
\usepackage{verbatim}
\usetikzlibrary{arrows}
\usetikzlibrary{shapes,snakes}
\usetikzlibrary{decorations.pathmorphing} 
\usetikzlibrary{fit}					
\usetikzlibrary{backgrounds}	

\usepackage{microtype}
\usepackage{balance}
\usepackage{setspace}
\graphicspath{{./}{./graphics/}}
\newcolumntype{H}{>{\setbox0=\hbox\bgroup}c<{\egroup}@{}}

\newcolumntype{R}[1]{>{\RaggedLeft\arraybackslash}} 
\newcolumntype{L}[1]{>{\RaggedRight\arraybackslash}}

\newcommand{\eg}{\emph{e.g.}}
\newcommand{\ie}{\emph{i.e.}}

\newtheorem{property}{\hspace{-1em}\bfseries{Property}}
\newtheorem{lem}{\hspace{-1em}\bfseries{Lemma}}
\newtheorem{Definition}{\bfseries{Definition}}

\usepackage{microtype}

\providecommand{\mat}[1]{\boldsymbol{\mathrm{#1}}}
\renewcommand{\vec}[1]{\boldsymbol{\mathrm{#1}}}

\DeclareMathOperator*{\argmax}{argmax}

\DeclareMathOperator{\hugeE}{\mbox{\huge\raise-0.3ex\hbox{E}}}
\DeclareMathOperator{\p}{\mathbb{P}}
\DeclareMathOperator{\hugep}{\mbox{\huge\raise-0.3ex\hbox{$\p$}}}

\providecommand{\mA}{\ensuremath{\mat{A}}}

\providecommand{\vc}{\ensuremath{\vec{c}}}

\usepackage{comment}

\usepackage{nicefrac}
\usepackage{algorithm}
\usepackage{algorithmicx}
\usepackage{algpseudocode}
\algblockdefx[parallel]{ParFor}{EndPar}[1][]{$\textbf{parallel for}$ #1 $\textbf{do}$}{$\textbf{end parallel}$}
\algrenewcommand{\alglinenumber}[1]{\fontsize{6.5}{7}\selectfont#1}
\algtext*{EndPar}

\algblockdefx[parallel]{parfor}{endpar}[1][]{$\textbf{parallel for}$ #1 $\textbf{do}$}{$\textbf{end parallel}$}
\algrenewcommand{\alglinenumber}[1]{\scriptsize#1:}

\algblockdefx[parallel]{parallelfor}{parallelend}
[1][]{\textbf{parallel for} #1}
{\textbf{end parallel}}

\usepackage{nicefrac}

\algnotext{EndFor}
\algnotext{EndIf}
\algnotext{EndProcedure}

\begin{document}

\title{Linear-time Hierarchical Community Detection}

\author{Ryan A. Rossi}
\orcid{1234-5678-9012-3456}
\affiliation{
\institution{Adobe Research}
}

\author{Nesreen K. Ahmed}
\affiliation{
\institution{Intel Labs}
}

\author{Eunyee Koh}
\affiliation{
\institution{Adobe Research}
}

\author{Sungchul Kim}
\affiliation{
\institution{Adobe Research}
}
\email{}

\renewcommand\shortauthors{Rossi, R.~A. et al.}

\begin{abstract}
Community detection in graphs has many important and fundamental applications including in distributed systems, compression, image segmentation, divide-and-conquer graph algorithms such as nested dissection, document and word clustering, circuit design, among many others. Finding these densely connected regions of graphs remains an important and challenging problem. Most work has focused on scaling up existing methods to handle large graphs. These methods often partition the graph into two or more communities. In this work, we focus on the problem of \emph{hierarchical community detection} (\ie, finding a hierarchy of dense community structures going from the lowest granularity to the largest) and describe an approach that runs in linear time with respect to the number of edges and thus fast and efficient for large-scale networks. The experiments demonstrate the effectiveness of the approach quantitatively. Finally, we show an application of it for visualizing large networks with hundreds of thousands of nodes/links.
\end{abstract}

\ccsdesc[500]{Mathematics of computing~Graph algorithms}
\ccsdesc[500]{Mathematics of computing~Approximation algorithms}
\ccsdesc[500]{Mathematics of computing~Graph theory}
\ccsdesc[500]{Information systems~Data mining}
\ccsdesc[500]{Theory of computation~Graph algorithms analysis}
\ccsdesc[500]{Networks~Network types}

\keywords{Community detection, hierarchical communities, linear-time algorithms, 
label propagation,
graph clustering, graph mining
}

\maketitle

\section{Introduction} \label{sec:intro}
Communities of a graph are sets of nodes that are densely connected and close to one another in the graph~\cite{Fortunato2010}.
Communities are important for understanding complex systems modeled as graphs~\cite{graph-clustering-survey,Fortunato2010}.
In our modern age of big data, it has become increasingly important to study and understand complex systems that arise from large data of diversely interconnected entities such as biological networks \cite{alon2003biological}, social networks \cite{girvan2002community}, citation networks \cite{giles2006future}, among many others. 
Community detection in graphs has been one of the most fundamental tools for analyzing and understanding the components of complex networks and has been used for many real-world applications. It has been used extensively in applications to distributed systems~\cite{hendrickson1995improved, simon1991partitioning, van1995improved}, compression~\cite{rossi2015pmc-sisc, buehrer2008scalable}, image segmentation~\cite{shi2000normalized, felzenszwalb2004efficient}, document and word clustering \cite{dhillon2001co}, among others.

Communities are sets of vertices $C_1,\ldots,C_k$
such that each set $C_k$ has with more connections inside the set than outside~\cite{Fortunato2010}.
While there are many different methods for finding communities~\cite{graph-clustering-survey,Fortunato2010}, it is generally agreed that a community $C_k \subseteq V$ is ``good" if the induced subgraph is dense (\eg, many edges between the vertices in $C_k$) and there are relatively few edges from $C_k$ to other vertices $\bar{C_k} = V \setminus C_k$~\cite{graph-clustering-survey}.
Let $E(C_k)$ denote the set of edges between vertices in $C_k$ (internal edges) and $E(C_k,\bar{C}_k)$ be the set of all edges between $C_k$ and $\bar{C}_k$ (external edges).
Another desired property of a community $C_k$ is that vertices in $C_k$ are all close to one another, \ie, the distance between any two vertices $v,w \in C_k$ denoted as $\mathtt{dist}(v,w)$ is as small as possible (small proximity, distance).
Community detection aims to cut a graph into two or more sparsely interconnected dense subgraphs~\cite{Fortunato2010}.
Semantically, these subgraphs may represent a tightly-knit group of friends, a household or organization, web pages of the same general topic, or a group of researchers that frequently publish together.
In this work, we address the following problem:

\begin{Definition}[Hierarchical Community Detection]\label{def:hier-comm-detection}
Given an (un)directed graph $G=(V,E)$, the problem of hierarchical community detection is to find 
\begin{compactenum}[$\bullet$ \leftmargin=0.0em]
\item[(i)] 
a hierarchy of communities denoted as 
$\mathbb{H} = \{\mathbfcal{C}^1,\ldots,\mathbfcal{C}^L\}$
where $\mathbfcal{C}^{t}=\{C^{t}_1,\ldots,C^{t}_k\}$ are the communities at level $t$ in the hierarchy
\begin{align}
V_{t} = \bigcup_k C^{t}_k 
\;\;\text{ and }\;\;
|\mathbfcal{C}^1| < \cdots < |\mathbfcal{C}^t| < \cdots < |\mathbfcal{C}^L|
\end{align}
\item[(ii)] a hierarchy of 
community (super) graphs 
$G_1, \ldots, G_t, \ldots, G_L$ 
where $G_t=(V_t,E_t)$ succinctly captures the relationships between the communities (nodes in $G_t$) at a lower $t-1$ level in the hierarchy.
The hierarchy of community (super) graphs indicate how the functional units (communities) of the graph interact at each level and how they combine to form larger communities.
\end{compactenum}
\end{Definition}

\begin{figure*}[h!]
\vspace{-1mm}
\centering
\subfigure[$\mathbfcal{C}^{1}$ (40 communities)]{
\includegraphics[width=0.32\linewidth]{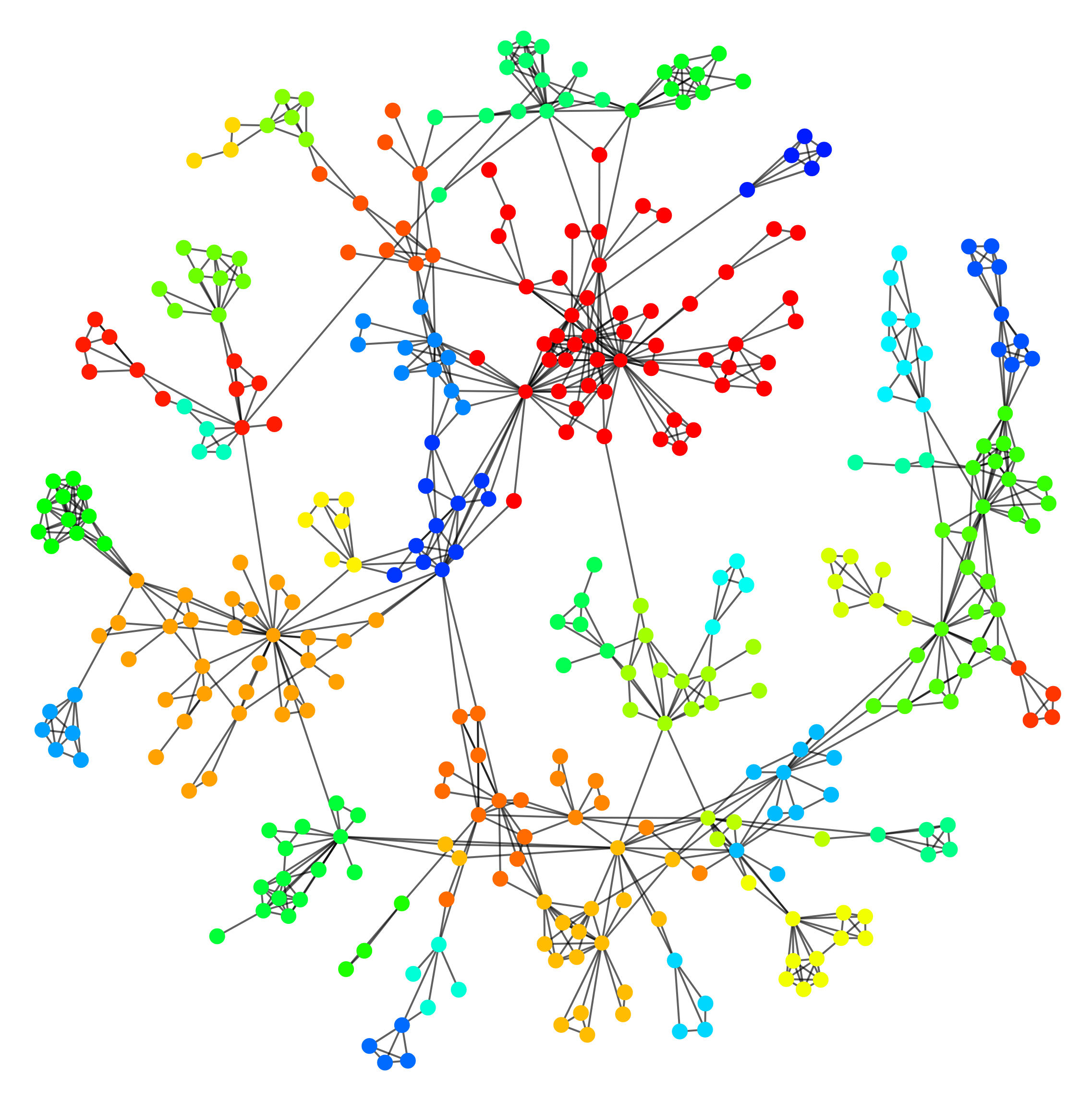}\label{fig:LP-vs-hLP-netscience-traditional-LP}}
\hfill
\subfigure[$\mathbfcal{C}^{2}$ (6 communities)]{
\includegraphics[width=0.32\linewidth]{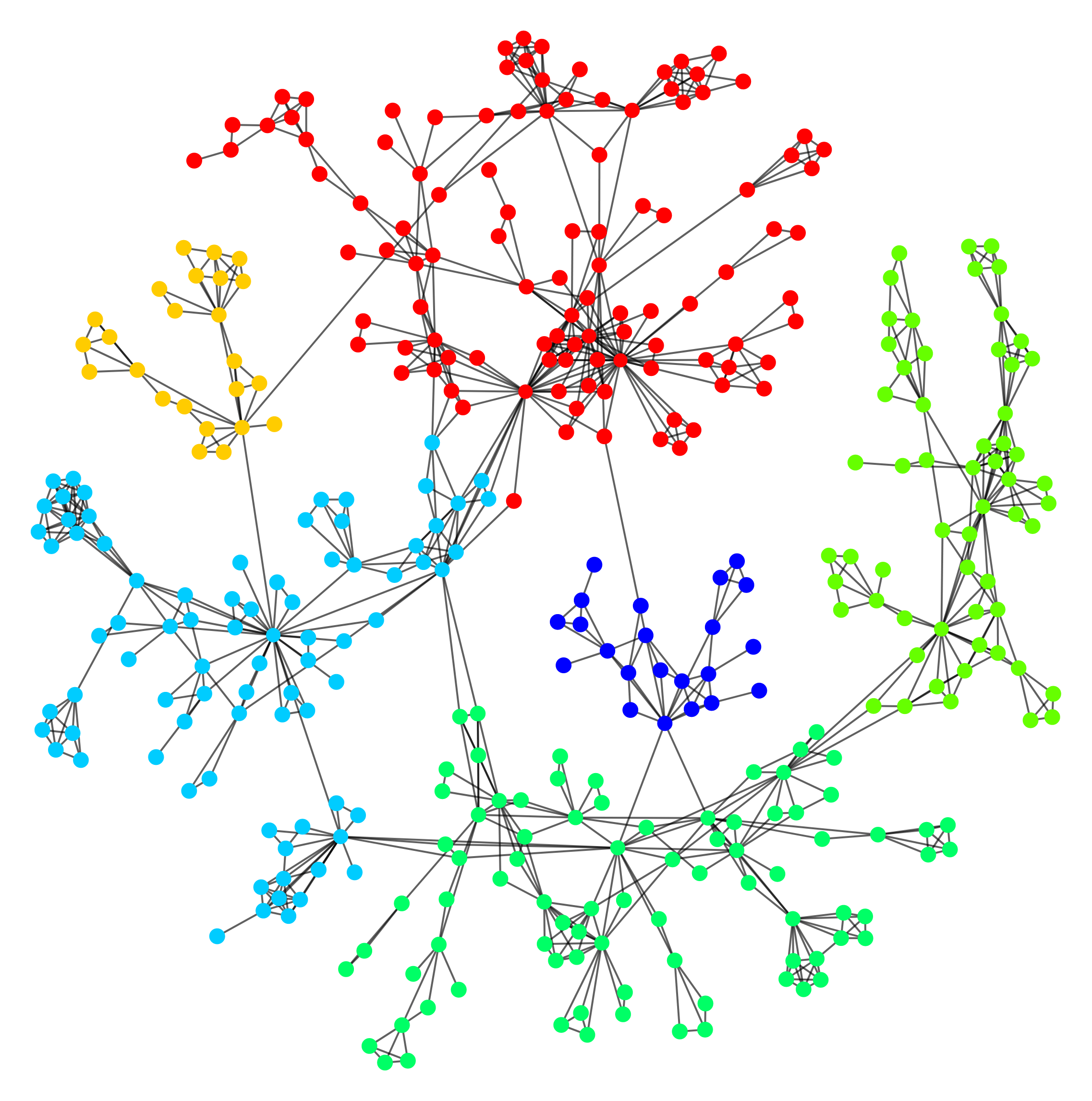}\label{fig:LP-vs-hLP-netscience-L2}}
\hfill
\subfigure[$\mathbfcal{C}^{3}$ (2 communities)]{
\includegraphics[width=0.32\linewidth]{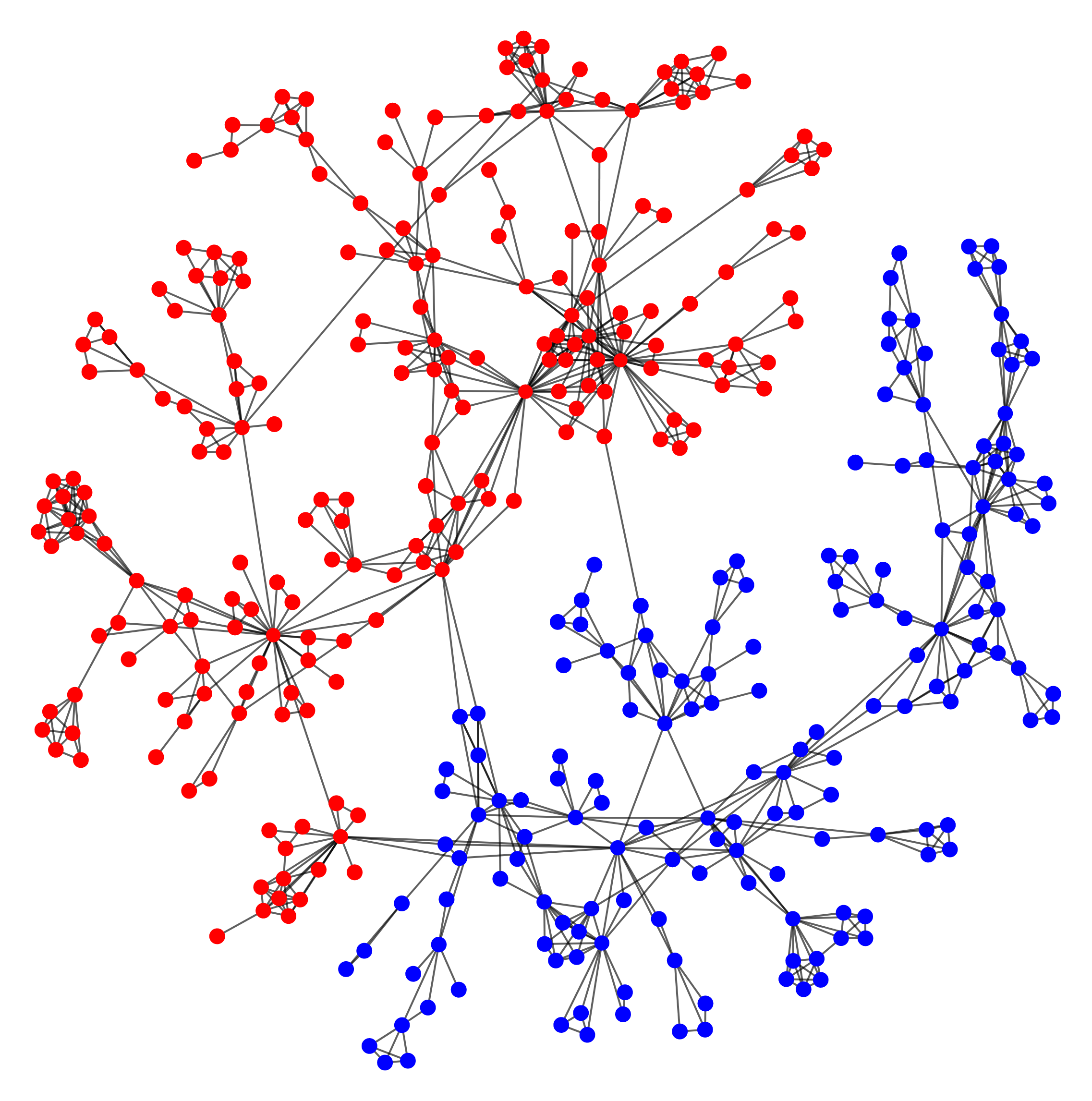}\label{fig:LP-vs-hLP-netscience-L3}}
\vspace{-3.5mm}
\caption{Network science co-authorship network.
\textsc{h}LP summarizes the higher-order organization of the network at multiple granularities as shown in \ref{fig:LP-vs-hLP-netscience-traditional-LP}-\ref{fig:LP-vs-hLP-netscience-L3}. 
Node color encodes community assignment.
See text for discussion.
}
\label{fig:LP-vs-hLP-netscience}
\vspace{-1mm}
\end{figure*}

While there have been a lot of work on community detection~\cite{graph-clustering-survey,Fortunato2010}, most research
(i) does not address the hierarchical community detection problem (Definition~\ref{def:hier-comm-detection}) \emph{or} are
(ii) inefficient for large networks with a worst-case time (and space) complexity that is \emph{not} linear in the number of edges.
In this work, we describe an approach called \textsc{hLP} that addresses both these limitations.
In particular, \textsc{h}LP solves the hierarchical community detection problem by detecting a hierarchy of communities (going from the lowest to highest granularity) along with a hierarchy of community (super) graphs that reveal the higher-order organization and components at each level and how these components interact with one another to form larger components at a higher-level in the hierarchy.
Most importantly, \textsc{h}LP is fast and efficient for large networks with a worst-case time complexity that is linear in the number of edges whereas the space complexity of \textsc{h}LP is linear in the number of nodes.

{
\algblockdefx[parallel]{parfor}{endpar}[1][]{$\textbf{parallel for}$ #1 $\textbf{do}$}{$\textbf{end parallel}$}
\algrenewcommand{\alglinenumber}[1]{\fontsize{7.0}{8.0}\selectfont#1\;\;}
\begin{figure}[b!]
\vspace{-1mm}
\begin{center}
\begin{algorithm}[H]
\caption{\,
Hierarchical Community Detection ({\textsc{h}LP})
}
\label{alg:hier-LP}
\begin{spacing}{1.15}
\small
\begin{algorithmic}[1]
\smallskip
\Require a graph $G=(V,E)$ 
\Ensure hierarchical communities $\mathbb{H} = \{\mathbfcal{C}^1,\ldots,\mathbfcal{C}^L\}$ 
\smallskip
\State Set $G_{0} \leftarrow G$ to be the initial graph and $t \leftarrow 1$
\Repeat
\State $\mathbfcal{C}^{t}\leftarrow$ \textsc{LabelProp}($G_{t-1}$) 
\State $G_{t}=(V_{t},E_{t}) \leftarrow \,$ \textsc{CreateSuperGraph}($G_{t-1}$, $\mathbfcal{C}^{t}$) via Eq.~\ref{eq:super-edges-for-t}
\State $t \leftarrow t + 1$
\smallskip
\Until{$|V_{t}|<2$}
\label{algline:check-convergence} 
\Comment{Stop when no nodes to combine}
\end{algorithmic}
\end{spacing}
\end{algorithm}
\end{center}
\vspace{-4mm}
\end{figure}
}

{
\algblockdefx[parallel]{parfor}{endpar}[1][]{$\textbf{parallel for}$ #1 $\textbf{do}$}{$\textbf{end parallel}$}
\algrenewcommand{\alglinenumber}[1]{\fontsize{7}{8}\selectfont#1\;\;}
\begin{figure}[b!]
\vspace{-4mm}
\begin{center}
\begin{algorithm}[H]
\caption{\, Create Super Graph}
\label{alg:create-supergraph}
\begin{spacing}{1.15}
\small
\begin{algorithmic}[1]
\smallskip
\Require a graph $G_{t-1} = (V_{t-1}, E_{t-1})$, communities $\mathbfcal{C}^{t}$ from $G_{t-1}$
\Ensure community (super) graph $G_t=(V_t,E_t)$ for layer $t$ 
\smallskip
\State $V_{t} \leftarrow \mathbfcal{C}^{t-1}$ where $\mathbfcal{C}^{t-1} = \{C_1, \ldots, C_k\}$ \Comment{Super node set}
\State $E_{t} \leftarrow \emptyset$ \Comment{Super edge set}

\State Let $\vc$ be the community assignment vector where $c_i=k$ if $v_i \in C_k$
\parfor[$i \in V_{t-1}$]
\For{$j \in \Gamma_{i}$} \Comment{Neighbor of vertex $i$}
\If{$c_i \not= c_j$ \textbf{and} $(c_i,c_j) \not\in E_{t}$}
\State $E_{t} \leftarrow E_{t} \cup (c_i,c_j)$
\EndIf
\EndFor
\endpar
\end{algorithmic}
\end{spacing}
\vspace{-0.mm}
\end{algorithm}
\end{center}
\vspace{-4mm}
\end{figure}
}

\section{Approach} \label{sec:approach}
This section describes our fast linear-time approach for revealing hierarchical communities in large graphs.
Given $G$, the algorithm outputs a hierarchy of communities $\mathbb{H} = \{\mathbfcal{C}^{1}, \ldots, \mathbfcal{C}^{L}\}$ where $L$ is the number of layers (\ie, levels in the community hierarchy $\mathbb{H}$).
A summary of the approach is shown in Algorithm~\ref{alg:hier-LP}.
There are two general steps: Label Propagation (Section~\ref{sec:label-prop}) and 
Super Graph Construction 
(Section~\ref{sec:super-graph-construction}).

\subsection{Label Propagation}\label{sec:label-prop}
Note $\Gamma(v_i) = \{j \in V \,|\, (i,j) \in E\}$ is the set of neighbors of node $i$.
The first step performs label propagation.
In particular, the approach begins with each node belonging to its own community.
For each node $v_i \in V$ (or edge), we assign it to the community $C_k \in \mathbfcal{C}$ that has the maximum number of neighbors $\Gamma(v_i)$ in it.
More formally, 
\begin{equation}\label{eq:comm-detection-single-node-obj}
\argmax_{C_k \in \,\mathbfcal{C}} \sum_{v_j \in\, \Gamma(v_i)} \mathbb{I} \big[\, v_j \in C_k \,\big]
\end{equation}\noindent
where for any predicate $p$ the indicator function $\mathbb{I}[ p ] = 1$ iff $p$ holds and $0$ otherwise.
Hence, $\mathbb{I} \big[\, v_j \in C_k \,\big]=1$ iff $v_j \in C_k$, and 0 otherwise. 
In other words, every node $v_i \in V$ is assigned the label that appears the most frequent in the 1-hop neighborhood of the node
Eq.~\ref{eq:comm-detection-single-node-obj} can be easily replaced/modified to take into account other important aspects.
The algorithm converges when an iteration results in no further changes (\ie, no new assignments are made) or if the max number of iterations is reached which can be interactively tuned by the user.
Upon each iteration, we compute a random permutation and use this ordering to assign nodes (or edges) to communities.
To further speedup the approach, we leverage the number of previous iterations that the community assignment of a node (or edge) remained unchanged (\ie, the community of $v_i$ remained stable over the last $t$ iterations).
In particular, let $\delta$ denote a hyperparameter that controls the number of previous iterations that the community assignment of a node or edge must remain unchanged before it is declared as final.
Thus, each iteration of the approach can be defined over the set $S$ of graph elements (nodes/edges) that are still active, \ie, $T_i < \delta$ where $T_i$ denotes the number of subsequent iterations that $v_i$ has remained unchanged (\emph{w.r.t.} community assignment).
Fast and efficient localized updates are performed when new nodes/edges arrive. 

\subsection{Super Graph Construction}\label{sec:super-graph-construction}
Given a graph $G_{t-1}$ and $\mathbfcal{C}^{t} = \{C_1^{t}, \ldots, C_k^{t}\}$, Algorithm~\ref{alg:create-supergraph} computes the community (super) graph $G_{t}=(V_{t},E_{t})$ for layer $t$ in the community hierarchy where $V_{t} \leftarrow \mathbfcal{C}^{t}$
and thus the number of nodes in $G_{t}$ 
is $n_{t} = |\mathbfcal{C}^{t}|$, \ie, the number of communities detected in the previous graph $G_{t-1}$ (or level in the community hierarchy).
Similarly, an edge $(i,j) \in E_{t}$ iff there is an edge between $C_i^{t}$ and $C_j^{t}$ in $G_{t-1}$, 
\ie, a link exists between a node $r \in V_{t-1}$ assigned to community $C_i^{t}$ and another node $s \in V_{t-1}$ assigned to community $C_j^{t}$.
More formally,
\begin{equation}\label{eq:super-edges-for-t}
E_{t} = \big\lbrace (i,j) \,:\,  r \in C_i^{t}, \,s \in C_j^{t} \wedge\; (r,s) \in E_{t-1} \,\wedge\; i\not=j \big\rbrace
\end{equation}\noindent

\begin{property}
Let $E_t(C_i, C_j)$ be the set of edges between $C_i$ and $C_j$ (cut set), 
then the number of edges $|E_{t+1}|$ in the next level $t+1$ is:
\begin{align}
|E_{t+1}| = \sum_{C_i \in \mathbfcal{C}^{t}} \;\;\sum_{C_j \in \mathbfcal{C}^{t}} |E_t(C_i, C_j)| \;\;\text{ s.t. }\; i<j
\end{align}\noindent
Note $|E_t(C_i, C_j)|$ does not include multi-edges.
\end{property}

Algorithm~\ref{alg:create-supergraph} returns $G_t=(V_t,E_t)$ for layer $t$ in the hierarchy. 
The approach terminates when $|V_{t}|<2$ as shown in Algorithm~\ref{alg:hier-LP}.
Hence, \textsc{h}LP terminates when there are no nodes remaining to combine.

\begin{property}\label{prop:node-and-edge-size-shrink}
Let $|E(G_t)|$ and $|V(G_t)|$ be the number of edges and nodes in $G_t$ and $G_{0} \leftarrow G$, then
\begin{align}
|E(G_0)| \geq \cdots \geq |E(G_L)| \;\;\text{ and }\;\;
|V(G_0)| \geq \cdots \geq |V(G_L)|
\end{align}
\end{property}\noindent
Property~\ref{prop:node-and-edge-size-shrink} has a number of important and useful implications that are leveraged in Section~\ref{sec:analysis}.

\section{Analysis} \label{sec:analysis}
This section shows the worst-case time and space complexity of the proposed approach.
Let $L$ denote the number of layers (hierarchies) and let $T$ denote the maximum number of iterations at any given layer.
Both $L$ and $T$ are small.
Further, let $N = |V|$ denote the number of nodes and let $M = |E|$ denote the number of edges in $G$.

\subsection{Time Complexity}\label{sec:time-complexity}
\begin{lem}\label{lem:time-complexity}
The worst-case time complexity of \emph{hierarchical label propagation} is
\begin{equation}
\mathcal{O}(LTM) = \mathcal{O}(M)
\end{equation}\noindent
where $L$ and $T$ are small constants.
Therefore, the time complexity is linear in the number of edges $M$ in the graph.
\end{lem}

\smallskip\noindent\textbf{Supergraph construction}:
The worst-case time complexity of Algorithm~\ref{alg:create-supergraph} is $\mathcal{O}(|E_{t-1}|)$.
This is bounded above by the number of edges denoted as $|E|$ in the input graph $G$.

\subsection{Space Complexity}\label{sec:space-complexity}
\begin{lem}\label{lem:space-complexity}
The space complexity of hierarchical label propagation is
\begin{equation}
\mathcal{O}(NL)
\end{equation}\noindent
where $L$ is a small constant.
Therefore, the space complexity is linear in the number of nodes in $G$.
\end{lem}
Lemma~\ref{lem:space-complexity} assumes the node community assignments at each layer are stored and given as output to the user.
However, this information can be significantly compressed by storing only the community assignments at the first layer, and then storing only how these communities are merged at each subsequent layer.

\smallskip\noindent\textbf{Supergraph construction}:
The worst-case space complexity of Algorithm~\ref{alg:create-supergraph} is $\mathcal{O}(|E_{t-1}|)$.
Similar to time complexity, this is bounded above by the number of edges denoted as $|E|$ in the input graph $G$.

\begin{figure*}[h!]
\vspace{-1mm}
\centering
\subfigure[road-luxembourg ($\mathbfcal{C}^{2}$)] 
{
\includegraphics[width=0.32\linewidth]{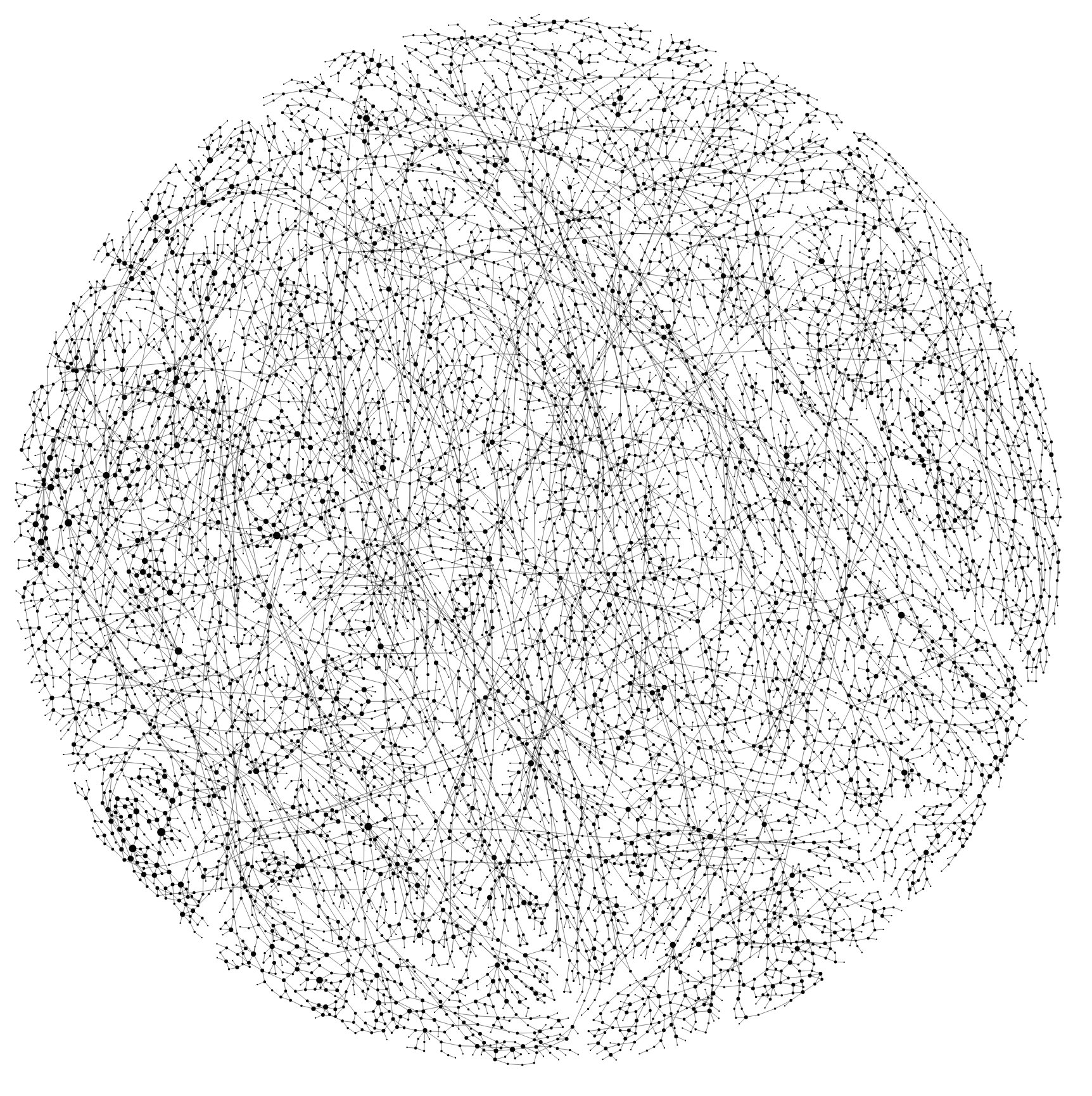}
\label{fig:supergraph-road-luxembourg-layer1}}
\hfill
\subfigure[$\mathbfcal{C}^{3}$]
{
\includegraphics[width=0.32\linewidth]{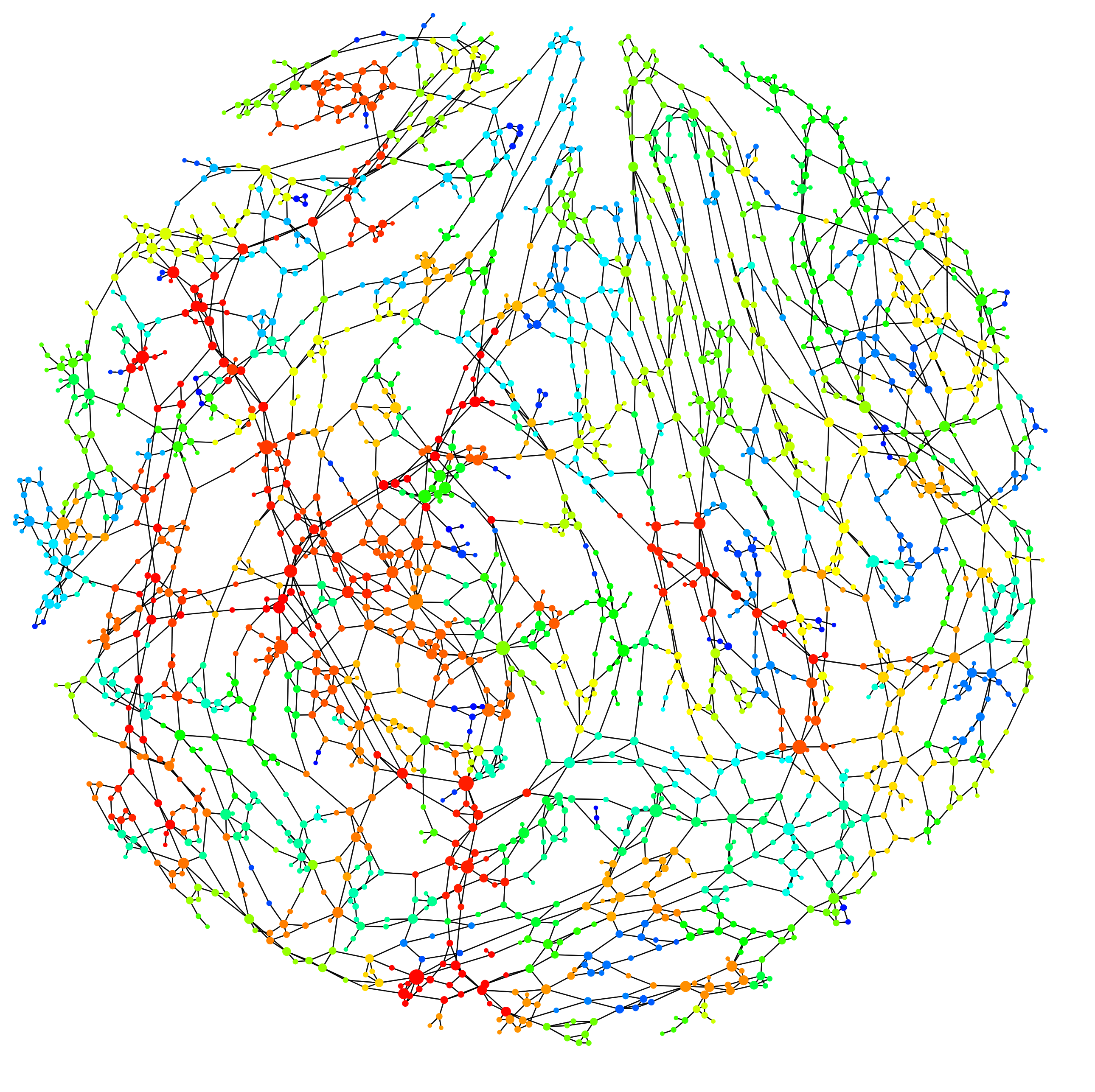}
\label{fig:supergraph-road-luxembourg-layer2}}
\hfill
\begin{minipage}[b]{0.32\textwidth}
\centering
\vspace{-4mm}
\subfigure[$\mathbfcal{C}^{4}$]
{
\includegraphics[width=0.85\linewidth]{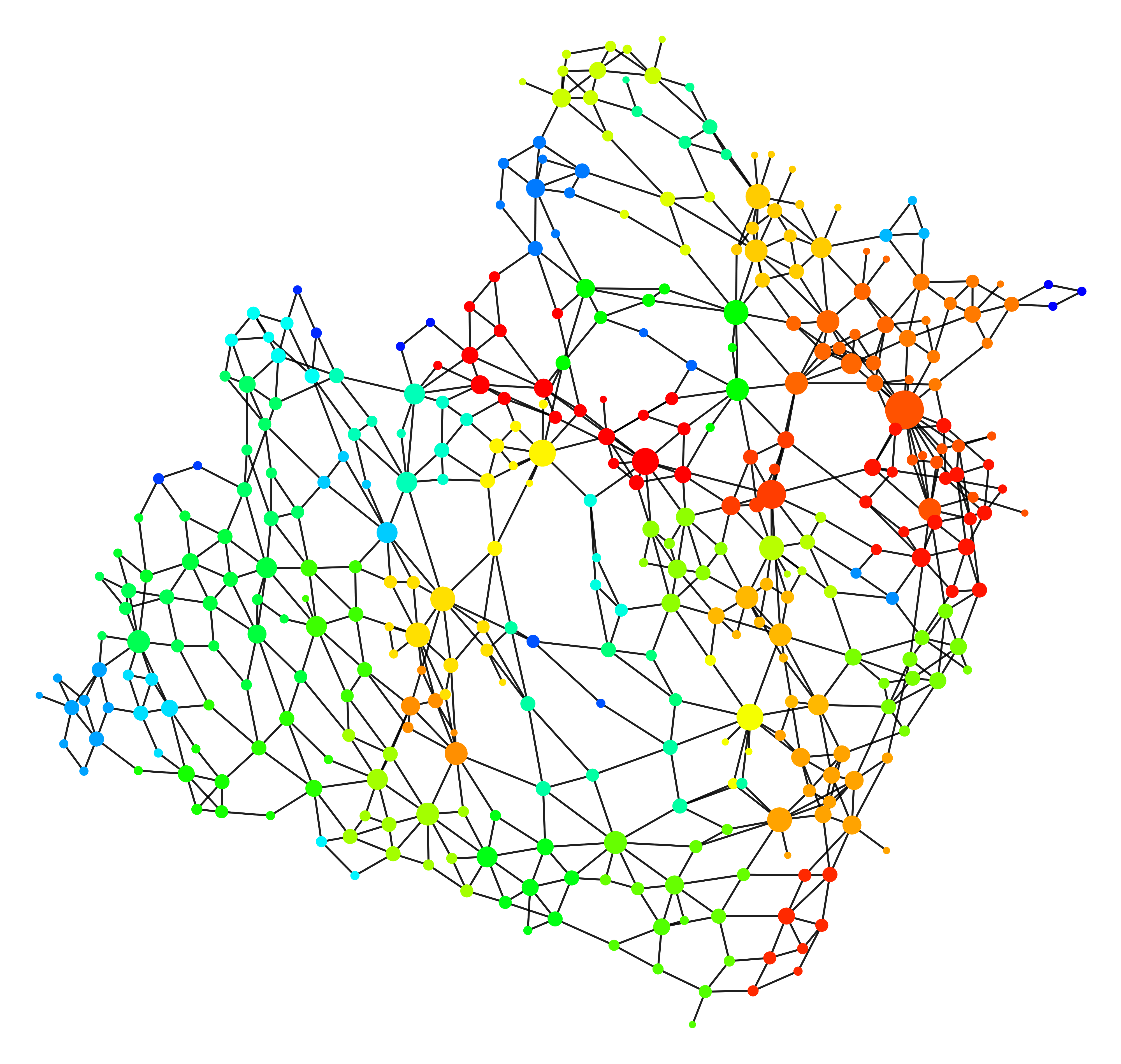}
}

\vspace{-2mm}
\subfigure[$\mathbfcal{C}^{5}$]
{
\includegraphics[width=0.55\linewidth]{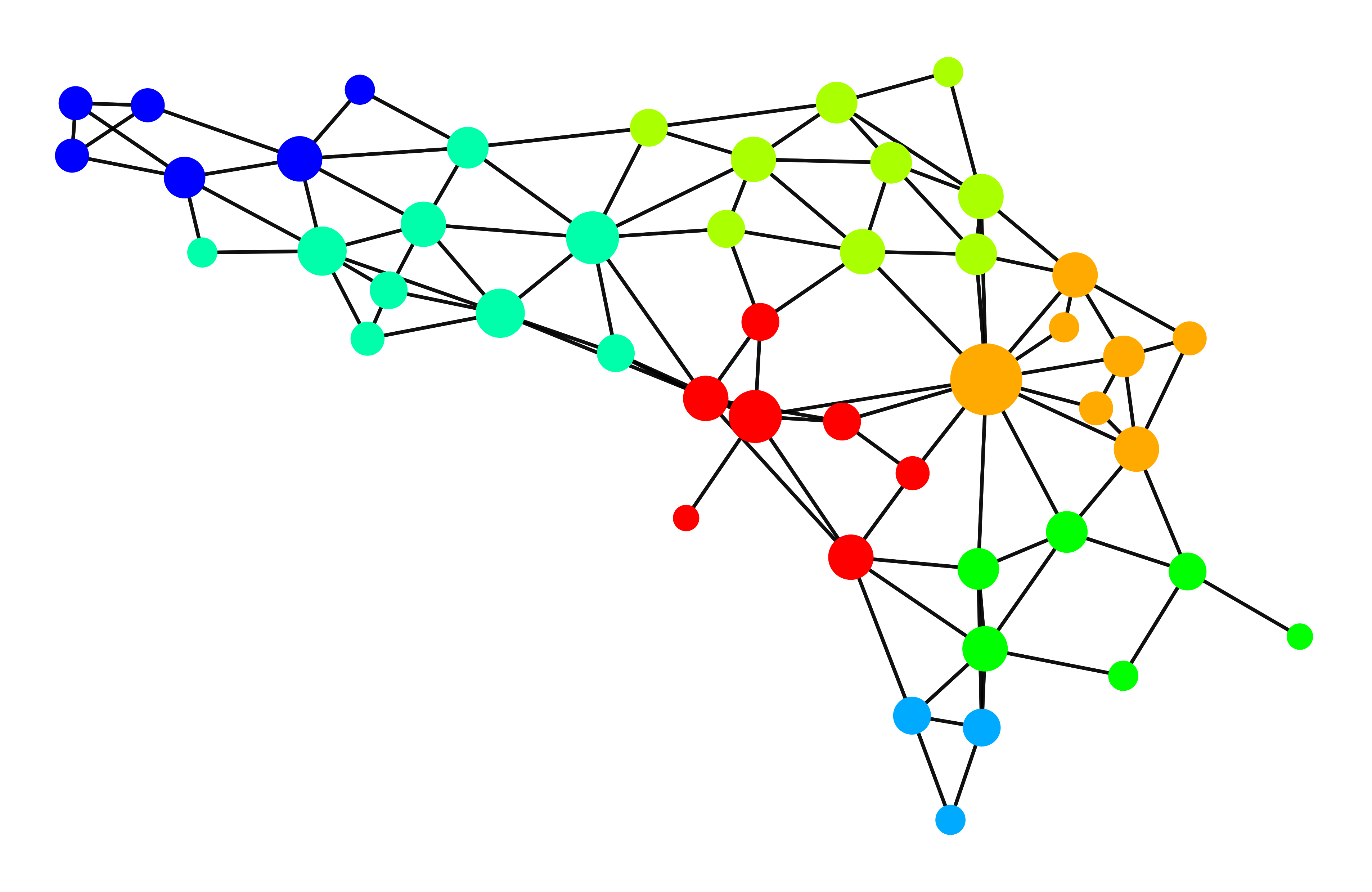}
}
\end{minipage}

\vspace{-3.5mm}
\caption{
In this case study, the network data is a road network of luxembourg consisting of 114,600 nodes and 239,332 edges making it impossible to visualize the entire network.
There are 9,452 communities in $\mathbfcal{C}^{2}$ and therefore impossible to visualize by assigning each community a unique color.
(a) Super graph derived after first layer consisting of 9,452 supernodes (communities) with 25,386 superedges (between community edges).
(b) consists of 2,023 communities with only 6,588 between community edges whereas (c)-(d) consists of 372 and 48 communities with 1,580 and 214 between community edges, respectively.
Nodes are weighted by degree. 
See text for discussion.
}
\vspace{-1.0mm}
\label{fig:supergraph-road-luxembourg-new}
\end{figure*}

\section{Experiments} \label{sec:exp}
The experiments in this section are designed to investigate the quality of the communities revealed by \textsc{h}LP and the utility of the hierarchical communities for a visualization application.
For comparison, we use a wide variety of graphs from different application domains including social networks (soc), biological/protein networks (bio), infrastructure networks (inf), web graphs (web), road networks (road), and collaboration networks (ca).
Due to space constraints, network statistics were removed but can be accessed online at \url{http://networkrepository.com} along with the data~\cite{nr}.

\subsection{Comparison}

\subsubsection{Baseline methods.}
For fair comparison, we use baselines that are fast with \emph{linear-time} complexity (with the exception of Louvain):
{\smallskip
\begin{compactenum}[$\bullet$ \leftmargin=0em]
\item \textbf{Densest Subgraph} (\textbf{DS})~\cite{khuller2009finding}:
This method finds an approximation of the densest subgraph in $G$ using degeneracy ordering,
and removes this subgraph. This is repeated until all nodes have been assigned.

\item \textbf{KCore Communities} (\textbf{KCore})~\cite{rossi2015pmc-sisc,shin2016corescope}:
Many have observed that the largest k-core subgraphs of a real-world network are highly dense subgraphs that often contain the max clique~\cite{rossi2015pmc-sisc}.
The KCore baseline simply uses the maximum k-core subgraph as $S$ and $\bar{S} = V \setminus S$.

\item \textbf{Label Propagation} (\textbf{LP})~\cite{raghavan2007near}:
Label propagation takes a labeling of the graph, then for each node, the label is updated according to the label that occurs the most among its neighbors.
This is repeated until convergence.

\item \textbf{Louvain} (\textbf{Louv})~\cite{blondel2008fast}:
Louvain performs a greedy optimization of modularity by forming small, locally optimal communities then grouping each community into one node. This two-phase process is repeated 
until modularity cannot be maximized locally. 

\item \textbf{Spectral Clustering}  (\textbf{Spec})~\cite{chung1997spectralbook}:
This baseline uses spectral clustering on the normalized Laplacian of the adjacency matrix to greedily build the sweeping cluster that minimizes conductance. 

\end{compactenum}
\smallskip}
\noindent

\begin{table}[h!]
\centering
\setlength{\tabcolsep}{3.5pt}
\renewcommand{\arraystretch}{1.1} 
\caption{
Quantitative evaluation of the methods (modularity).
The best result from each graph is bold.
Note \textsc{h}LP is the proposed method.
}
\vspace{-3mm}
\label{table:quant-mod}
\small
\begin{tabularx}{1.0\linewidth}{lHrr rr rH HrHHHHH}
\toprule
&& 
\multicolumn{1}{l}{\rotatebox{0}{\textbf{DS}}} &
\multicolumn{1}{l}{\rotatebox{0}{\textbf{KCore}}} &
\multicolumn{1}{l}{\rotatebox{0}{\textbf{LP}}} &
\multicolumn{1}{l}{\rotatebox{0}{\textbf{Louv}}} &
\multicolumn{1}{l}{\rotatebox{0}{\textbf{Spec}}} &&&
\multicolumn{1}{l}{\rotatebox{0}{\textbf{\textsc{h}LP}}} & 
\\
\midrule
\textsf{soc-yahoo-msg} && 0.0003 & 0.0004 & 0.0479 & 0.0394 & 0.0005 &&& \textbf{0.0569} \\ 
\textsf{bio-gene} && 0.0195 & 0.0217 & 0.0315 & 0.0408 & -0.0208 &&& \textbf{0.0846} \\ 
\textsf{ca-cora} && 0.0089 & 0.0304 & 0.0444 & 0.0608 & 0.0164 &&& \textbf{0.1026} \\ 
\textsf{soc-terror} &&  0.0888 & 0.0892 & 0.0967 & 0.0967 & 0.0999 &&& \textbf{0.1243} \\ 
\textsf{inf-US-powerGrid} && 0.0027 & 0.0027 & 0.0061 & 0.0212 & 0.1127 &&& \textbf{0.1242} & \\
\textsf{web-google} && 0.0272 & 0.0275 & 0.0429 & 0.0471 & 0.1010 &&& \textbf{0.1122} & \\ 
\textsf{ca-CSphd} && 0.0224 & 0.0224 & 0.0234 & 0.0198 & 0.0131 &&& \textbf{0.1201} & \\
\textsf{ca-netscience} && 0.0164 & 0.0168 & 0.1063 & 0.0561 & 0.1229 &&& \textbf{0.1233} & \\
\textsf{road-luxem.} && 0.0629 & 0.0629 & 0.0077 & 0.0046 & -0.1170 &&& \textbf{0.1141} & \\
\textsf{bio-DD21} && 0.0865 & 0.0866 & 0.0106 & 0.0202 & 0.1241 &&& \textbf{0.1247} & \\ 
\bottomrule
\end{tabularx}
\end{table}

\subsubsection{Quantitative evaluation.}
We quantitatively evaluate the communities using modularity~\cite{newman2001structure}.
Modularity is defined as:
\begin{equation} \label{eq:modularity}
\mathbb{E(\vc)} = \frac{1}{2m} \sum_{ij} \Bigg[ A_{ij} - \frac{d_i d_j}{2m} \Bigg] \delta(c_i, c_j)
\end{equation}\noindent
where $M$ is the number of edges, $\mA$ is the adjacency matrix with $A_{ij}=1$ if $(i,j) \in E$ and 0 otherwise; $d_i$ and $d_j$ is the degree of node $i$ and $j$; $c_i$ and $c_j$ are the community assignments of node $i$ and node $j$; and $\delta$ is an indicator function such that $\delta(c_i, c_j)=1$ if $c_i=c_j$ and $0$ otherwise.
We report the best result from any layer/level in the community hierarchy.
Results are provided in Table~\ref{table:quant-mod}.
Notably, \textsc{h}LP outperforms all the other baseline methods across all graphs as shown in Table~\ref{table:quant-mod}.
\textsc{h}LP reveals better high quality communities across a wide variety graphs from different application domains (social, biological, infrastructure, among others) as shown in Table~\ref{table:quant-mod}.
Overall, \textsc{h}LP typically achieves at least an order of magnitude improvement over the other baseline methods.

Now we investigate the communities found by \textsc{h}LP by overlaying the community assignments on top of the network structure (node-link diagram).
The communities given by \textsc{h}LP at different levels in the hierarchy are shown in Figure~\ref{fig:LP-vs-hLP-netscience} for the network science co-authorship network.
Communities in \ref{fig:LP-vs-hLP-netscience-traditional-LP} represent small groups of researchers that frequently publish together whereas communities in \ref{fig:LP-vs-hLP-netscience-L2} represent different research areas and so on.

\subsubsection{Runtime Performance} \label{sec:exp-runtime-perf}
Figure~\ref{fig:supergraph-road-luxembourg-new} visualizes the important components (functional modules) of a large road network from luxembourg at multiple scales (layers).
Note that using a serial python implementation of the proposed method takes only 10.2 seconds to derive the initial 9,452 communities visualized in Figure~\ref{fig:supergraph-road-luxembourg-layer1}.
However, the next layer is orders of magnitude faster taking less than a second
(0.611 sec.) and the runtime steadily decreases as a function of the supergraph size (number of supernodes, superedges) and the number of iterations to converge in the preceding layers. 
Furthermore, the number of iterations until convergence also steadily decreases as the number of layers increases.

\subsection{Visualizing Large Networks}
One important application of \textsc{h}LP is visualization of large networks.
In Figure~\ref{fig:supergraph-road-luxembourg-new}, we use \textsc{h}LP to compute a hierarchy of communities for a large real-world network consisting of 114,600 nodes and 239,332 edges.
While it is impractical and often impossible to visualize such a large network, we can use \textsc{h}LP to summarize the graph structure at multiple levels as shown in Figure~\ref{fig:supergraph-road-luxembourg-new}.
Instead of visualizing the graph at the level of intersections (nodes in the original road network), we can instead visualize the graph at a higher-level where nodes represent something more meaningful, \eg, instead of intersections, nodes at layer 2 shown in Figure~\ref{fig:supergraph-road-luxembourg-layer2} might represent neighborhoods and edges represent routes from one neighborhood to another.
Thus, \textsc{h}LP uncovers the \emph{hierarchical higher-order organization} of complex networks.

\bibliographystyle{ACM-Reference-Format}
\bibliography{paper}


\begin{thebibliography}{00}


\ifx \showCODEN    \undefined \def \showCODEN     #1{\unskip}     \fi
\ifx \showDOI      \undefined \def \showDOI       #1{#1}\fi
\ifx \showISBNx    \undefined \def \showISBNx     #1{\unskip}     \fi
\ifx \showISBNxiii \undefined \def \showISBNxiii  #1{\unskip}     \fi
\ifx \showISSN     \undefined \def \showISSN      #1{\unskip}     \fi
\ifx \showLCCN     \undefined \def \showLCCN      #1{\unskip}     \fi
\ifx \shownote     \undefined \def \shownote      #1{#1}          \fi
\ifx \showarticletitle \undefined \def \showarticletitle #1{#1}   \fi
\ifx \showURL      \undefined \def \showURL       {\relax}        \fi
\providecommand\bibfield[2]{#2}
\providecommand\bibinfo[2]{#2}
\providecommand\natexlab[1]{#1}
\providecommand\showeprint[2][]{arXiv:#2}

\bibitem[\protect\citeauthoryear{Alon}{Alon}{2003}]%
        {alon2003biological}
\bibfield{author}{\bibinfo{person}{Uri Alon}.} \bibinfo{year}{2003}\natexlab{}.
\newblock \showarticletitle{Biological networks: the tinkerer as an engineer}.
\newblock \bibinfo{journal}{{\em Science\/}} \bibinfo{volume}{301},
  \bibinfo{number}{5641} (\bibinfo{year}{2003}), \bibinfo{pages}{1866--1867}.
\newblock


\bibitem[\protect\citeauthoryear{Blondel, Guillaume, Lambiotte, and
  Lefebvre}{Blondel et~al\mbox{.}}{2008}]%
        {blondel2008fast}
\bibfield{author}{\bibinfo{person}{Vincent~D Blondel},
  \bibinfo{person}{Jean-Loup Guillaume}, \bibinfo{person}{Renaud Lambiotte},
  {and} \bibinfo{person}{Etienne Lefebvre}.} \bibinfo{year}{2008}\natexlab{}.
\newblock \showarticletitle{Fast unfolding of communities in large networks}.
\newblock \bibinfo{journal}{{\em JSTAT\/}} \bibinfo{number}{10}
  (\bibinfo{year}{2008}).
\newblock


\bibitem[\protect\citeauthoryear{Buehrer and Chellapilla}{Buehrer and
  Chellapilla}{2008}]%
        {buehrer2008scalable}
\bibfield{author}{\bibinfo{person}{Gregory Buehrer} {and}
  \bibinfo{person}{Kumar Chellapilla}.} \bibinfo{year}{2008}\natexlab{}.
\newblock \showarticletitle{A scalable pattern mining approach to web graph
  compression with communities}. In \bibinfo{booktitle}{{\em WSDM}}.
  \bibinfo{pages}{95--106}.
\newblock


\bibitem[\protect\citeauthoryear{Chung}{Chung}{1997}]%
        {chung1997spectralbook}
\bibfield{author}{\bibinfo{person}{Fan~RK Chung}.}
  \bibinfo{year}{1997}\natexlab{}.
\newblock \bibinfo{booktitle}{{\em Spectral graph theory}}.
\newblock \bibinfo{publisher}{AMS}.
\newblock


\bibitem[\protect\citeauthoryear{Dhillon}{Dhillon}{2001}]%
        {dhillon2001co}
\bibfield{author}{\bibinfo{person}{Inderjit~S Dhillon}.}
  \bibinfo{year}{2001}\natexlab{}.
\newblock \showarticletitle{Co-clustering documents and words using bipartite
  spectral graph partitioning}. In \bibinfo{booktitle}{{\em SIGKDD}}.
\newblock


\bibitem[\protect\citeauthoryear{Felzenszwalb and Huttenlocher}{Felzenszwalb
  and Huttenlocher}{2004}]%
        {felzenszwalb2004efficient}
\bibfield{author}{\bibinfo{person}{Pedro~F Felzenszwalb} {and}
  \bibinfo{person}{Daniel~P Huttenlocher}.} \bibinfo{year}{2004}\natexlab{}.
\newblock \showarticletitle{Efficient graph-based image segmentation}.
\newblock \bibinfo{journal}{{\em IJCV\/}} \bibinfo{volume}{59},
  \bibinfo{number}{2} (\bibinfo{year}{2004}).
\newblock


\bibitem[\protect\citeauthoryear{Fortunato}{Fortunato}{2010}]%
        {Fortunato2010}
\bibfield{author}{\bibinfo{person}{Santo Fortunato}.}
  \bibinfo{year}{2010}\natexlab{}.
\newblock \showarticletitle{{Community detection in graphs}}.
\newblock \bibinfo{journal}{{\em Phy. Rep.\/}} \bibinfo{number}{3}
  (\bibinfo{year}{2010}).
\newblock
\showISSN{0370-1573}


\bibitem[\protect\citeauthoryear{Giles}{Giles}{2006}]%
        {giles2006future}
\bibfield{author}{\bibinfo{person}{C~Lee Giles}.}
  \bibinfo{year}{2006}\natexlab{}.
\newblock \showarticletitle{The future of citeseer: citeseer x}. In
  \bibinfo{booktitle}{{\em ECML}}. Springer, \bibinfo{pages}{2--2}.
\newblock


\bibitem[\protect\citeauthoryear{Girvan and Newman}{Girvan and Newman}{2002}]%
        {girvan2002community}
\bibfield{author}{\bibinfo{person}{Michelle Girvan} {and}
  \bibinfo{person}{Mark~EJ Newman}.} \bibinfo{year}{2002}\natexlab{}.
\newblock \showarticletitle{Community structure in social and biological
  networks}.
\newblock \bibinfo{journal}{{\em PNAS\/}} \bibinfo{volume}{99},
  \bibinfo{number}{12} (\bibinfo{year}{2002}), \bibinfo{pages}{7821--7826}.
\newblock


\bibitem[\protect\citeauthoryear{Hendrickson and Leland}{Hendrickson and
  Leland}{1995}]%
        {hendrickson1995improved}
\bibfield{author}{\bibinfo{person}{Bruce Hendrickson} {and}
  \bibinfo{person}{Robert Leland}.} \bibinfo{year}{1995}\natexlab{}.
\newblock \showarticletitle{An improved spectral graph partitioning algorithm
  for mapping parallel computations}.
\newblock \bibinfo{journal}{{\em SIAM SISC\/}} \bibinfo{volume}{16},
  \bibinfo{number}{2} (\bibinfo{year}{1995}).
\newblock


\bibitem[\protect\citeauthoryear{Khuller and Saha}{Khuller and Saha}{2009}]%
        {khuller2009finding}
\bibfield{author}{\bibinfo{person}{Samir Khuller} {and} \bibinfo{person}{Barna
  Saha}.} \bibinfo{year}{2009}\natexlab{}.
\newblock \showarticletitle{On finding dense subgraphs}. In
  \bibinfo{booktitle}{{\em ICALP}}.
\newblock


\bibitem[\protect\citeauthoryear{Newman}{Newman}{2001}]%
        {newman2001structure}
\bibfield{author}{\bibinfo{person}{M.E.J. Newman}.}
  \bibinfo{year}{2001}\natexlab{}.
\newblock \showarticletitle{{The structure of scientific collaboration
  networks}}.
\newblock \bibinfo{journal}{{\em PNAS\/}} \bibinfo{volume}{98},
  \bibinfo{number}{2} (\bibinfo{year}{2001}), \bibinfo{pages}{404}.
\newblock


\bibitem[\protect\citeauthoryear{Raghavan, Albert, and Kumara}{Raghavan
  et~al\mbox{.}}{2007}]%
        {raghavan2007near}
\bibfield{author}{\bibinfo{person}{Usha~Nandini Raghavan},
  \bibinfo{person}{R{\'e}ka Albert}, {and} \bibinfo{person}{Soundar Kumara}.}
  \bibinfo{year}{2007}\natexlab{}.
\newblock \showarticletitle{Near linear time algorithm to detect community
  structures in large-scale networks}.
\newblock \bibinfo{journal}{{\em Physical Review E\/}} \bibinfo{volume}{76},
  \bibinfo{number}{3} (\bibinfo{year}{2007}), \bibinfo{pages}{036106}.
\newblock


\bibitem[\protect\citeauthoryear{Rossi and Ahmed}{Rossi and Ahmed}{2016}]%
        {nr}
\bibfield{author}{\bibinfo{person}{Ryan~A. Rossi} {and}
  \bibinfo{person}{Nesreen~K. Ahmed}.} \bibinfo{year}{2016}\natexlab{}.
\newblock \showarticletitle{An Interactive Data Repository with Visual
  Analytics}.
\newblock \bibinfo{journal}{{\em SIGKDD Exp.\/}} (\bibinfo{year}{2016}).
\newblock
\showURL{%
\url{http://networkrepository.com}}


\bibitem[\protect\citeauthoryear{Rossi, Gleich, and Gebremedhin}{Rossi
  et~al\mbox{.}}{2015}]%
        {rossi2015pmc-sisc}
\bibfield{author}{\bibinfo{person}{Ryan~A. Rossi}, \bibinfo{person}{David
  Gleich}, {and} \bibinfo{person}{Assefaw Gebremedhin}.}
  \bibinfo{year}{2015}\natexlab{}.
\newblock \showarticletitle{Parallel Maximum Clique Algorithms with
  Applications to Network Analysis}.
\newblock \bibinfo{journal}{{\em SISC\/}} (\bibinfo{year}{2015}).
\newblock


\bibitem[\protect\citeauthoryear{Schaeffer}{Schaeffer}{2007}]%
        {graph-clustering-survey}
\bibfield{author}{\bibinfo{person}{Satu~Elisa Schaeffer}.}
  \bibinfo{year}{2007}\natexlab{}.
\newblock \showarticletitle{Graph clustering}.
\newblock \bibinfo{journal}{{\em Comp. sci. rev.\/}} \bibinfo{volume}{1},
  \bibinfo{number}{1} (\bibinfo{year}{2007}), \bibinfo{pages}{27--64}.
\newblock


\bibitem[\protect\citeauthoryear{Shi and Malik}{Shi and Malik}{2000}]%
        {shi2000normalized}
\bibfield{author}{\bibinfo{person}{Jianbo Shi} {and} \bibinfo{person}{Jitendra
  Malik}.} \bibinfo{year}{2000}\natexlab{}.
\newblock \showarticletitle{Normalized cuts and image segmentation}.
\newblock \bibinfo{journal}{{\em TPAMI\/}} \bibinfo{volume}{22},
  \bibinfo{number}{8} (\bibinfo{year}{2000}), \bibinfo{pages}{888--905}.
\newblock


\bibitem[\protect\citeauthoryear{Shin, Eliassi-Rad, and Faloutsos}{Shin
  et~al\mbox{.}}{2016}]%
        {shin2016corescope}
\bibfield{author}{\bibinfo{person}{Kijung Shin}, \bibinfo{person}{Tina
  Eliassi-Rad}, {and} \bibinfo{person}{Christos Faloutsos}.}
  \bibinfo{year}{2016}\natexlab{}.
\newblock \showarticletitle{{CoreScope:} Graph Mining Using k-Core
  Analysis--Patterns, Anomalies and Algorithms}. In \bibinfo{booktitle}{{\em
  ICDM}}.
\newblock


\bibitem[\protect\citeauthoryear{Simon}{Simon}{1991}]%
        {simon1991partitioning}
\bibfield{author}{\bibinfo{person}{Horst~D Simon}.}
  \bibinfo{year}{1991}\natexlab{}.
\newblock \showarticletitle{Partitioning of unstructured problems for parallel
  processing}.
\newblock \bibinfo{journal}{{\em Comp. Sys. in Eng.\/}} \bibinfo{volume}{2},
  \bibinfo{number}{2} (\bibinfo{year}{1991}).
\newblock


\bibitem[\protect\citeauthoryear{Van~Driessche and Roose}{Van~Driessche and
  Roose}{1995}]%
        {van1995improved}
\bibfield{author}{\bibinfo{person}{Rafael Van~Driessche} {and}
  \bibinfo{person}{Dirk Roose}.} \bibinfo{year}{1995}\natexlab{}.
\newblock \showarticletitle{An improved spectral bisection algorithm and its
  application to dynamic load balancing}.
\newblock \bibinfo{journal}{{\em Parallel comp.\/}} (\bibinfo{year}{1995}).
\newblock


\end{thebibliography}

\end{document}